# Adaptive Adjustment of Noise Covariance in Kalman Filter for Dynamic State Estimation

Shahrokh Akhlaghi, *Student Member, IEEE*
Ning Zhou, *Senior Member, IEEE*
Electrical and Computer Engineering Department,
Binghamton University, State University of New York,
Binghamton, NY 13902, USA
{sakhlag1, ningzhou}@binghamton.edu

Zhenyu Huang, *Senior Member, IEEE*
Pacific Northwest National Laboratory,
Richland, WA 99352, USA
zhenyu.huang@pnnl.gov

*Abstract*—Accurate estimation of the dynamic states of a synchronous machine (e.g., rotor's angle and speed) is essential in monitoring and controlling transient stability of a power system. It is well known that the covariance matrixes of process noise (*Q*) and measurement noise (*R*) have a significant impact on the Kalman filter's performance in estimating dynamic states. The conventional *ad-hoc* approaches for estimating the covariance matrixes are not adequate in achieving the best filtering performance. To address this problem, this paper proposes an adaptive filtering approach to adaptively estimate *Q* and *R* based on *innovation* and *residual* to improve the dynamic state estimation accuracy of the extended Kalman filter (EKF). It is shown through the simulation on the two-area model that the proposed estimation method is more robust against the initial errors in *Q* and *R* than the conventional method in estimating the dynamic states of a synchronous machine.

## I. INTRODUCTION

Timely and accurately estimating the dynamic states of a synchronous machine (e.g., rotor angle and rotor speed) is important for monitoring and controlling the transient stability of a power system over wide areas [1]. With the worldwide deployment of phasor measurement units (PMUs), many research efforts have been made to estimate the dynamic states and improve the estimation accuracy using PMU data [2]–[11], among which the Kalman filtering (KF) techniques play an essential role. For instance, Huang et al. [2] proposed an extended Kalman filtering (EKF) approach to estimate the dynamic states using PMU data. Ghahremani and Innocent [3] proposed the EKF with unknown inputs to simultaneously estimate dynamic states of a synchronous machine and unknown inputs. [4]-[7] proposed the unscented Kalman filtering to estimate power system dynamic states. Zhou et al. [8] proposed an ensemble Kalman filter approach to simultaneously estimate the dynamic states and parameters. Akhlaghi, Zhou and Huang [9]-[10] proposed an adaptive interpolation approach to mitigate the impact of non-linearity in dynamic state estimation (DSE). These studies have laid a solid ground for estimating the dynamic states of a power system and also revealed some needs for further studies.

One important problem that needs to be addressed in using the KF is how to properly set up the covariance matrixes of process noise (i.e., *Q*) and measurement noise (i.e., *R*). Note that the performance of the KF is highly affected by *Q* and *R* [12]. Improper choice of *Q* and *R* may significantly degrade the KF's performance and even make the filter diverge [13]. To determine *Q* and *R*, almost all the previous DSE studies used an ad-hoc procedure, in which *Q* and *R* are assumed to be constant during the estimation and are manually adjusted by trial-and-error approaches. Note that because the noise levels may change for different applications and users of DSE can have different backgrounds, it can be very challenging to use such an ad-hoc approach to properly set up *Q* and *R*.

To address this challenge, this paper proposes an estimation approach to adaptively adjust *Q* and *R* at each step of the EKF to improve DSE accuracy. An *innovation*-based method is used to adaptively adjust *Q*. A *residual*-based method is used to adaptively adjust the *R*. A simple example is used to evaluate the impact of *Q* and *R* on the performance of EKF. Then, performance of the proposed approach is evaluated using a two-area model [1].

The rest of paper is organized as follows: Section II reviews the dynamic model of a synchronous machine used for DSE. In Section III, the adaptive EKF approach is proposed. Sections IV and V present a case study and simulation results. Conclusions are drawn in Section VI.

## II. DYNAMIC STATE ESTIMATION MODEL

This section gives a brief review on the dynamic model of a synchronous machine to be used by the EKF for DSE. The 4$^{th}$ order differential equations of a synchronous machine in a local *d-q* reference frame is given by (1). (Readers may refer to [9]-[11] for more details):

$$\begin{cases} \dfrac{d\delta}{dt} = \omega_0 \Delta\omega_r & (1.a) \\[4pt] \dfrac{d\Delta\omega_r}{dt} = \dfrac{1}{2H}(T_m - T_e - K_D \Delta\omega_r) & (1.b) \\[4pt] \dfrac{de'_q}{dt} = \dfrac{1}{T'_{d0}}\left(E_{fd} - e'_q - (x_d - x'_d)i_d\right) & (1.c) \\[4pt] \dfrac{de'_d}{dt} = \dfrac{1}{T'_{q0}}\left(-e'_d + (x_q - x'_q)i_q\right) & (1.d) \end{cases}$$

In (1), the 4 states, $\delta$, $\omega_r$, $e'_d$ and $e'_q$, are the rotor angles in radians, rotor speeds in per-unit (*pu*) and transient voltages in *pu* along *d* and *q* axes, respectively. $\omega_0 = 2\pi f_0$ is the synchronous speed; $T_m$ and $T_e$ are the mechanical and the electric air-gap torque in *pu*; and parameters $H$ and $K_D$ are the inertia and damping factor, respectively; $E_{fd}$ is the internal field voltage. Variables $x_d$ and $x_q$ are the synchronous reactance; $x'_d$ and $x'_q$ are the transient reactance along *d* and *q* axes, respectively. $i_d$ and $i_q$ are the stator currents along *d* and *q* axes, respectively. $T'_{d0}$ and $T'_{q0}$ are the open circuit time constants in the *dq0* frame.

To facilitate the notation for applying the EKF to DSE of a synchronous machine, (1) is transformed into a general discrete state space model as shown in (2) and (3) with sampling interval of $\Delta t$ using the modified Euler method [11].

$$\begin{cases} x_k = \Phi(x_{k-1}, u_{k-1}) + w_{k-1} \\ z_k = h(x_k, u_k) + v_k \end{cases} \quad (2.a)$$

$$\Phi^{[1]}_{k-1} \approx \left.\dfrac{\partial \Phi(x_{k-1}, u_{k-1})}{\partial x}\right|_{x_k = \hat{x}^+_{k-1}} , \quad H^{[1]}_k \approx \left.\dfrac{\partial h(x_k, u_k)}{\partial x}\right|_{x_k = \hat{x}^-_k} \quad (2.b)$$

$$x_k = \begin{bmatrix} \delta & \Delta\omega & e'_q & e'_d \end{bmatrix}^T$$
$$u_k = \begin{bmatrix} T_m & E_{fd} & i_R & i_I \end{bmatrix}^T \quad (3)$$
$$z_k = \begin{bmatrix} e_R & e_I \end{bmatrix}^T$$

Here, subscript *k* is the time index, which indicates the time instance at $k\Delta t$. Symbols $x_k$, $u_k$, and $z_k$ are the state, input and measurement output, respectively. Functions $\Phi(*)$ and $h(*)$ are the state transition and measurement function, respectively. $\Phi^{[1]}_{k-1}$ is the Jacobian matrix of the state transition matrix at step *k*-1, and $H^{[1]}_k$ is the Jacobian matrix of the measurement function at step *k*. In (2), vectors $w_k$ and $v_k$ are the state process noise and measurement noise, respectively. Their mean and variance are denoted by (4) [11]. Here, symbol $E(*)$ represents the expected value. Symbols $Q_k$ and $R_k$ are the covariance matrixes of process noise and measurement noise respectively at step *k*.

$$E(w_k) = 0 \quad , \quad E(w_k w_k^T) = Q_k \quad (4.a)$$
$$E(v_k) = 0 \quad , \quad E(v_k v_k^T) = R_k \quad (4.b)$$

## III. ADAPTIVE EXTENDED KALMAN FILTER APPROACH

This section describes the conventional extended Kalman filter (CEKF) and proposes an adaptive extended Kalman filter (AEKF) approach which adaptively estimates $Q_{k-1}$ and $R_k$.

### A. Conventional Extended Kalman Filter

The CEKF consists of the following 3 steps. Readers may refer to [11] for more details about the CEKF.

**Step (0) – Initialization:**

To initialize the CEKF, the mean values and covariance matrix of the states are set up at *k* = 0 as in (5).

$$\hat{x}^+_0 = E(x_0) \quad (5.a)$$
$$P^+_0 = E\left[(x_0 - \hat{x}^+_0)(x_0 - \hat{x}^+_0)^T\right] \quad (5.b)$$

where the superscript "+" indicates that the estimate is *a posteriori*, and *P* is the state covariance matrix.

**Step (I) – Prediction:**

The state and its covariance matrix at *k*-1 are projected one step forward to obtain the *a priori* estimates at *k* as in (6).

$$\underbrace{\hat{x}^-_k = \Phi(\hat{x}^+_{k-1}, u_{k-1})}_{\text{Predicted State Estimate}} \quad (6.a)$$

$$\underbrace{P^-_k = \Phi^{[1]}_{k-1} P^+_{k-1} \Phi^{[1]T}_{k-1} + Q_{k-1}}_{\text{Priori Covariance Matrix}} \quad (6.b)$$

**Step (II) - Correction:**

The actual measurement is compared with predicted measurement based on the *a priori* estimate. The difference is used to obtain an improved *a posteriori* estimate as in (7).

$$\underbrace{d_k = \left[z_k - h_k(\hat{x}^-_k)\right]}_{\text{Measurement innovation}} \quad (7.a)$$

$$\underbrace{S_k = H^{[1]}_k P^-_k H^{[1]T}_k + R_k}_{\text{Innovation Covariance}} \quad (7.b)$$

$$\underbrace{\bar{K}_k = P^-_k H^{[1]T}_k \left[S_k\right]^{-1}}_{\text{Kalman Gain}} \quad (7.c)$$

$$\underbrace{\hat{x}^+_k = \hat{x}^-_k + \bar{K}_k \left[d_k\right]}_{\text{Posteriori State Estimate}} \quad (7.d)$$

$$\underbrace{P^+_k = \left\{I - \bar{K}_k H^{[1]}_k\right\} P^-_k}_{\text{Posteriori Covariance Matrix}} \quad (7.e)$$

Note that to run the CEKF, users need to provide $Q_{k-1}$ in (6.b) and $R_k$ in (7.b). Performance of a CEKF depends on how well users can select the right $Q_{k-1}$ and $R_k$ for different applications. Conventionally, $R_k$ is often assigned as a constant matrix based on the instrument accuracy of the measurements. $Q_{k-1}$ is assigned as a constant matrix using a trial-and-error approach, which relies on users' experiences and background. As such, selection of $Q_{k-1}$ and $R_k$ is a challenge for the users of the CEKF.

### B. Adaptive Extended Kalman Filter (AEKF)

To address this challenge, this paper proposes an adaptive estimation approach to estimate $Q_{k-1}$ and $R_k$ in the EKF. Mehra [14] classified the adaptive estimation approaches into four categories: Bayesian, correlation, covariance matching and maximum likelihood approaches. The covariance matching is one of the well-known adaptive estimation approaches, which tunes the covariance matrix of the *innovation* or *residual* based on their theoretical values [15]. At the EKF's predication step, the *innovation* is the difference between the actual measurement and its predicted value, and it can be calculated by (7.a). On the other hand, the *residual* is the difference between actual measurement and its estimated value using the information available at step *k*, and it can be calculated by (8).

$$\underbrace{\varepsilon_k = \left[z_k - h_k(\hat{x}_k^+)\right]}_{\text{residual}} \quad (8)$$

Based on the above definitions, the $Q_{k-1}$ and $R_k$ can be estimated as the follows.

*1) Residual Based Adaptive Estimation of R*

The *innovation* based approach estimates the covariance matrix $R_k$ using (9) [12].

$$R_k = S_k - H_k^{[1]} P_k^- H_k^{[1]T} \quad (9)$$

Here $S_k$ is the covariance matrix of the *innovation*. Note that theoretically speaking, $R_k$ should be positive definite because it is a covariance matrix. Yet, its estimation equation (9) could not guarantee that the estimated $R_k$ be a positive definite matrix because the $R_k$ is estimated by subtracting the two positive definite matrixes. Therefore, to ensure a positive definite matrix, the *residual* based adaptive approach proposed by [16] is used by this paper to estimate $R_k$ using (10).

$$\begin{aligned}\overline{S}_k &= E\left[\varepsilon_k \varepsilon_k^T\right] = E\left[v_k v_k^T\right] - H_k^{[1]} P_k^- H_k^{[1]T} \\ R_k &= E\left[\varepsilon_k \varepsilon_k^T\right] + H_k^{[1]} P_k^- H_k^{[1]T}\end{aligned} \quad (10)$$

To implement (10), the expectation operation on $\varepsilon_k \varepsilon_k^T$ is approximated by averaging $\varepsilon_k \varepsilon_k^T$ over time. Instead of the using the moving window, this paper introduces a forgetting factor $0 < \alpha \leq 1$ in (11) to adaptively estimate $R_k$. Note that a larger $\alpha$ puts more weights on previous estimates and therefore incurs less fluctuation of $R_k$, and longer time delays to catch up with changes. This paper set $\alpha = 0.3$ for all the studies.

$$R_k = \alpha R_{k-1} + (1-\alpha)(\varepsilon_k \varepsilon_k^T + H_k^{[1]} P_k^- H_k^{[1]T}) \quad (11)$$

*2) Innovation Based Adaptive estimation of Q*

To adaptively estimate the $Q_{k-1}$, based on (2), the process noise can be calculated using (12).

$$w_{k-1} = x_k - \Phi(x_{k-1}, u_{k-1}) \quad (12)$$

From (6) and (7), it can be concluded that:

$$\begin{aligned}\hat{w}_{k-1} &= \hat{x}_k^+ - \Phi(\hat{x}_{k-1}^+, u_{k-1}) \\ &= x_k^+ - \hat{x}_k^- = \overline{K}_k \left[z_k - h_k(\hat{x}_k^-)\right] \\ &= \overline{K}_k d_k\end{aligned} \quad (13)$$

Therefore,

$$\begin{aligned}E\left[\hat{w}_{k-1}\hat{w}_{k-1}^T\right] &= E\left[\overline{K}_k (d_k d_k^T) \overline{K}_k^T\right] = \overline{K}_k E\left[d_k d_k^T\right] \overline{K}_k^T \\ \hat{Q}_{k-1} &= \overline{K}_k S \overline{K}_k^T\end{aligned} \quad (14)$$

Similar to the previous subsection, the paper uses a forgetting factor $\alpha$ to average estimates of *Q* over time as in (15).

$$Q_k = \alpha Q_{k-1} + (1-\alpha)(\overline{K}_k d_k d_k^T \overline{K}_k^T) \quad (15)$$

An implementation flowchart of the proposed AEKF algorithm is summarized in Fig. 1. Note that similar to the CEKF, users need to select the initial $Q_0$ and $R_0$ for AEKF in the initialization step. Different from the CEKF which keeps $Q_{k-1}$ and $R_k$ constant, the $Q_{k-1}$ and $R_k$ of the AEKF are adaptively estimated and updated during each correction step.

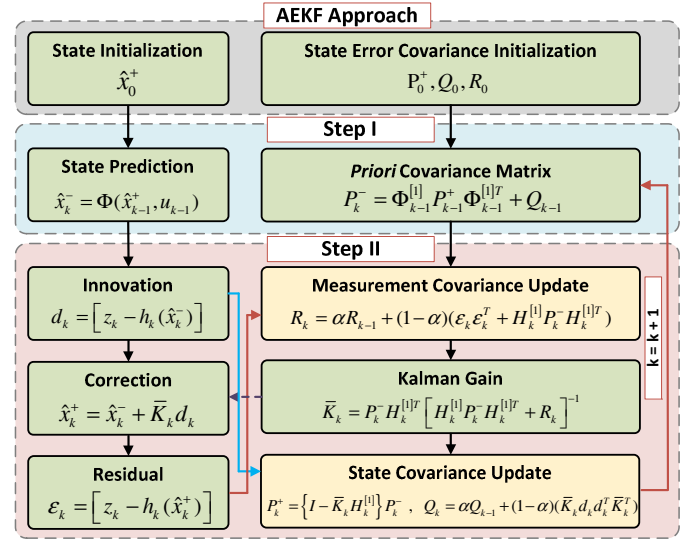

Fig. 1. Implementation flowchart of the proposed AEKF

## IV. CASE STUDY BASED ON A SIMPLE MODEL

In this section, a simple linear model described by (16) is used to compare the impact of the choice of $R_k$ and $Q_{k-1}$ on the performance of the CEKF and the AEKF. The simple and linear model with known noise features is used in this study to eliminate the potential impacts from non-linearity.

The model in (16) is a model of a vehicle tracking problem, which the vehicle is constrained to move in a straight line with a constant velocity. Let $p_k$ and $\dot{p}_k$ represent the vehicle position and velocity. The system state can be described by $x_k = [p_k, \dot{p}_k]$. It is assumed that sampled observations are acquired at discrete time interval $\Delta t$. The $w_{k-1}$ and $v_k$ are Gaussian white noise, whose variances are defined by (16.b).

$$\begin{cases}x_k = A x_{k-1} + w_{k-1} \\ z_k = H x_k + v_k\end{cases} \quad (16.a)$$

$$A = \begin{bmatrix}1 & \Delta t \\ 0 & 1\end{bmatrix}, \quad H = \begin{bmatrix}1 & 0\end{bmatrix}$$

$$Q_{true} = q_0 \begin{bmatrix}\Delta t^3/3 & \Delta t^2/2 \\ \Delta t^2/2 & \Delta t\end{bmatrix}, \quad R_{true} = r_0 [1] \quad (16.b)$$

The scalars $q_0$, $r_0$ and $\Delta t$ are set to be 0.01, 0.1 and 1, respectively. It is assumed that the vehicle starts from rest so that $x_0 = [0, 0]^T$. For testing the performance of the CEKF and AEKF, 100 time steps of simulation are generated using (16). To evaluate the impact of $Q_k$ and $R_k$ on the estimation accuracy, $x_0$ is set to its true values and $P_0$ is set to zeros to eliminate their impacts on the estimation. For the CEKF, $Q_k$ and $R_k$ are set by scaling $Q_{true}$ and $R_{true}$. As shown in Table I, the scaling factors are the multiples of 10. Using the same setup, the resulting mean squared errors (MSEs) of estimated position, i.e., $x(1)$, are summarized in Table I for the CEKF and in Table II for the AEKF.

It can be observed in Table I that the MSEs on the diagonal are same. Note that the ratio between $Q_k$ and $R_k$ are same for the diagonal elements. This observation indicates that it is the ratio between $Q_k$ and $R_k$ (instead of their individual values) that determines the performance of the CEKF. Also observe that the major diagonal, where $Q_k:R_k = Q_{true}:R_{true}$, have the smallest MSE (i.e. 0.051). The observation suggests that the optimal $Q/R$ ratio is around their true ratios. Also observe that the MSEs increase monotonously when the $Q/R$ ratio increases or decreases from its true value.

TABLE I. MSEs of the Estimated Position from the **CEKF**

| MSE | 0.01 $Q_{true}$ | 0.1 $Q_{true}$ | $Q_{true}$ | 10 $Q_{true}$ | 100 $Q_{true}$ |
|---|---|---|---|---|---|
| 0.01 $R_{true}$ | 0.051 | 0.083 | 0.0984 | 0.0987 | 0.0988 |
| 0.1 $R_{true}$ | 0.219 | 0.051 | 0.083 | 0.0984 | 0.0988 |
| $R_{true}$ | 3.54 | 0.219 | 0.051 | 0.083 | 0.098 |
| 10 $R_{true}$ | 27.28 | 3.54 | 0.219 | 0.051 | 0.083 |
| 100 $R_{true}$ | 41.40 | 27.28 | 3.54 | 0.219 | 0.051 |

TABLE II. MSE of the Estimated Position from the **AEKF**

| MSE | 0.01 $Q_{true}$ | 0.1 $Q_{true}$ | $Q_{true}$ | 10 $Q_{true}$ | 100 $Q_{true}$ |
|---|---|---|---|---|---|
| 0.01 $R_{true}$ | 0.0714 | 0.0787 | 0.0788 | 0.0788 | 0.0789 |
| 0.1 $R_{true}$ | 0.09 | 0.076 | 0.0783 | 0.0786 | 0.0787 |
| $R_{true}$ | 0.12 | 0.089 | 0.072 | 0.073 | 0.0736 |
| 10 $R_{true}$ | 0.13 | 0.089 | 0.087 | 0.076 | 0.076 |
| 100 $R_{true}$ | 0.17 | 0.089 | 0.081 | 0.078 | 0.074 |

Comparing Table I and Table II, one can observe that in general, the proposed AEKF produces smaller MSEs than the CEKF. The performance improvement of the AEKF is more significant when the MSEs of the CEKF are larger (at the bottom left of the table). The only exception to this improvement is at the major diagonal where the $Q_k:R_k= Q_{true}:R_{true}$, which is already an optimal $Q/R$ ratio setup for the CEKF. The MSEs for the AEKF is slightly larger than the CEKF. This may be because averaging operations in (10) and (11) are used to approximate the expectation operation, which will incur some estimation errors in $Q_k$ and $R_k$. Notice that in a real world application, the true value of $Q/R$ ratio and states are often not available and have to be estimated. The proposed AEKF provides a systematic way of estimating the Q/R ratios and can often achieve stable and smaller MSEs than most guessed values. Similar observations are made with the other state (i.e., x(2) speed) and are not presented here to be concise.

## V. Case Study Based on the Two-Area Model

To evaluate the performance of the proposed AEKF approach on the DSE of synchronous machines, the two-area four-machine system [1] shown in Fig. 2 is used to generate the simulation data. The simulation is performed using the Power System Toolbox (PST) [17]. A three-phase fault is applied to sending end of the line between buses 100 and 200 at 10.1 s. To reduce integration errors and capture the dynamics, the simulation time step is set to be 0.001 s.

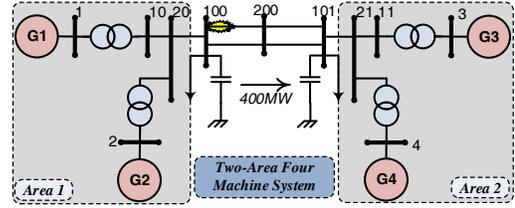

Fig. 2. The two-area four-machine system [1].

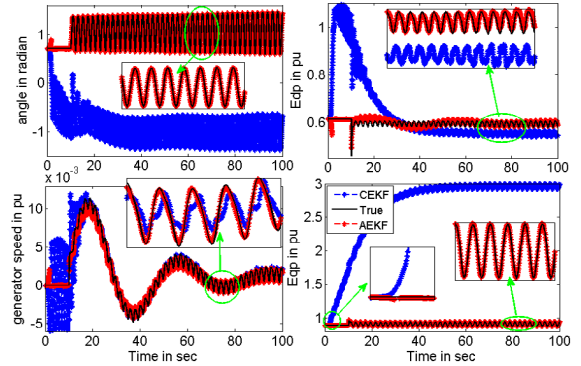

Fig. 3. Comparison of AEKF and CEKF when the initial $Q$ is set to be relatively less than the proper value.

It is assumed that all the generation buses are equipped with PMUs to measure the voltage phasors and current phasors in (3). To mimic the field measurements from the PMUs, the simulation data is decimated to 25 samples/s. And 4.0% noise in total vector errors [10] is added to the current and voltage phasors to consider the noise introduced by potential transformers and current transformers. Also, 4.0% noise is added to $E_{fd}$ and $T_m$.

Similar to section IV, in the initialization step, $x_0$ is set to its true values and $P_0$ is set to zeros to eliminate their impacts on the estimation. Assume that the $R_0$ is known based on the accuracy of measurement device and is set to be diag([0.04, 0.04])$^2$ to match the added measurement noise. The initial $Q_0$ is adjusted to set up the following four scenarios for comparing the estimation accuracy of the AEKF and CEKF.

**Scenario #1:** $Q_0$ is set to very small values (i.e. 1*e-08). The states estimated by the AEKF and CEKF are shown in Fig. 3 and their MSEs are summarized in Table III. Fig. 3 shows that with the same setup, the CEKF diverges while the AEKF converges. Table III shows that the MSEs of the AEKF is much smaller than those of the CEKF for all the estimated states. The observation indicates that when $Q_0$ is set up to be too small, the AEKF is robust against the improper setup and can estimate states accurately while the CEKF diverges.

**Scenario #2:** $Q_0$ is set to very large values (i.e. 1000). The states estimated by the AEKF and CEKF are shown in Fig. 4 and their MSEs are summarized in Table III. Fig. 4 shows that both the CEKF and AEKF converges and the states estimated by the AEKF stay closer to the true states than those estimated by the CEKF. Table III shows that the MSEs of the AEKF is smaller than those of the CEKF for all the estimated states. The observation indicates that when $Q_0$ is set up to be too large, both the AEKF and CEKF converge and the AEKF is more accurate than the CEKF, measured by MSEs.

**Scenario #3:** $Q_0$ is set to be close to the true values. As the true value of $Q_0$ is not accurately known, the final $Q$ resulting from the AEKF in Scenarios #2 is used. The MSEs of the estimated states are summarized in Table III. Table III shows that the MSEs of the AEKF is similar to those of the CEKF for all the estimated states. The observation indicates that when $Q_0$ is set up to be close to the true values, both the AEKF and CEKF converge and the AEKF has similar accuracy as the CEKF in the sense of MSEs.

**Scenario #4:** The setups of this scenarios are same as those for scenario #1 except that the Monte-Carlo simulation is used to generate $N = 200$ instances of simulation data with random noise. The estimated states are summarized in Fig. 5, which shows that observations made under scenarios #1 also apply to different noise instances.

TABLE III. COMPARISON OF THE MSEs OF THE ESTIMATED DYNAMIC STATES FROM THE CEKF AND AEKF

| Scenario # | | MSE | | | |
|---|---|---|---|---|---|
| | | $\delta$ | $\Delta\omega$ | $e'_d$ | $e'_q$ |
| 1 | CEKF | 3.77 | 1.21e-05 | 0.021 | 3.39 |
| | AEKF | 7.10e-05 | 1.25e-07 | 1.05e-04 | 3.03e-06 |
| 2 | CEKF | 1.74e-04 | 8.33e-07 | 4.43e-04 | 6.36e-05 |
| | AEKF | 2.53e-05 | 1.05e-07 | 9.74e-05 | 3.01e-06 |
| 3 | CEKF | 8.38e-05 | 3.98e-07 | 8.82e-05 | 3.38e-05 |
| | AEKF | 1.57e-05 | 1.09e-07 | 9.21e-05 | 2.78e-06 |

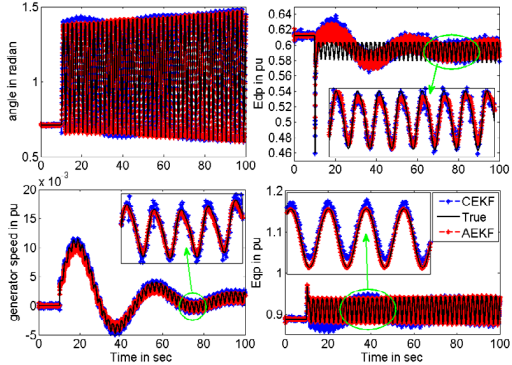

Fig. 4. Comparison of AEKF and CEKF when the initial $Q$ is set to be relatively greater than the proper value.

From the results of this section, it can be concluded that the proposed AEKF approch is robust against the initial errors in setting up the corvariance matrixes of process noise (i.e., $Q$) and measurement noise (i.e., $R$). The reason for scenarios #2 and #3 to have good performance is that we assume true $R$ and the measurements are ideal (meaning they match the model with no outliers, no losses). In this case, a large $Q$ would bias the EKF to believe the measurements, which would generate good estimate. If $R$ is unknown and/or measurements are not ideal, a blind selection of large $Q$ would fail to generate good estimates. We are testing $Q$ only as the first step to make it easier to show the effect of the AEKF. Scenario #1 is the most important case to examine. Future work will continue the research to test unknown $R$ and imperfect measurements.

## VI. CONCLUSIONS

This paper proposes an AEKF approach to adaptively estimate and adjust covariance matrixes $Q_{k-1}$ and $R_k$ for estimating the dynamic states of a synchronous machine. Also, it is shown through simulations using a simple model and the two-area system that the AEKF is more robust against the improper choice of initial $Q$ and $R$ than the CEKF. These simulation results suggest that the proposed AEKF can adaptively estimate $Q$ as well as $R$ and therefore relieve users' burden of choosing proper $Q$ and $R$ in the EKF.

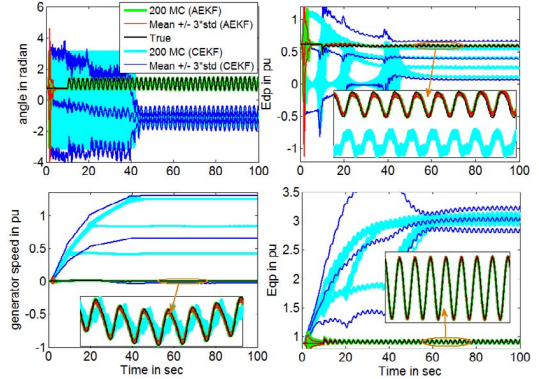

Fig. 5. Comparison of DSE results from the AEKF and CEKF approaches.